\documentclass[11pt,a4paper]{article}
\usepackage{epsfig}
\usepackage[nonstop]{feynmp}
\usepackage{amsmath}
\usepackage{cite}
\begin{document}

\title{{\Large \bf A Study of the Gluon Ladder in Diffractive Processes}}  
\author{{Jeppe~R.~Andersen\thanks{e-mail:
      andersen@hep.phy.cam.ac.uk}~~${}^{a}$, Agust{\'i}n Sabio~Vera\thanks{e-mail: sabio@mail.desy.de}~\thanks{Alexander von Humboldt Research Fellow}~~${}^{b}$}\\[4.5ex]  
$a~${\it Cavendish Laboratory, University of Cambridge, Madingley Road,}\\ 
{\it CB3 0HE Cambridge, UK}\\
$b~${\it II. Institut f{\"u}r Theoretische Physik, Universit{\"a}t 
Hamburg,}\\ {\it Luruper Chaussee 149, 22761~Hamburg, Germany} }

\maketitle

\vspace{-12cm}
\begin{flushright}  
Cavendish--HEP--2004/10\\
DESY 04--XXX\\
\end{flushright}  

\vspace{14cm}
\begin{abstract}  
  The solution to the non--forward BFKL equation in the Leading Logarithmic
  approximation is expressed in terms of a sum of iterations of its kernel
  directly in transverse momentum and rapidity space. Several studies of the
  non--forward solution are performed both at the level of the gluon Green's
  function and for a toy cross--section, including an analysis of the
  diffusion properties as found in this approach. The method developed in
  this paper allows for a direct inspection of the momenta in the BFKL ladder,
  and can be applied to solving the non--forward BFKL equation to
  next--to--leading logarithmic accuracy, when the corresponding kernel is
  available.
\end{abstract}  

\newpage

\section{Introduction}  

An interesting framework to study the behaviour of QCD scattering amplitudes 
in the limit of large centre--of--mass energies $\sqrt{s}$ and fixed momentum 
transfer $\sqrt{-t}$ is the Balitsky--Fadin--Kuraev--Lipatov (BFKL) 
formalism~\cite{LLBFKL}. 
The BFKL framework can be applied to processes characterised by a colour
octet exchange as well as colour singlet exchange (diffractive processes).
The case of $t=0$ colour singlet scattering corresponds to forward
scattering, whereas the case of $t\not=0$ is called non--forward scattering.
The optical theorem relates the amplitude for forward diffractive scattering
to the amplitude for a colour octet exchange, which will be exploited in the
check of some of the results derived in this paper. If the momentum transfer
is perturbative, i.e. $-t \gg \Lambda_{\rm QCD}^2$, it is possible to use the
non--forward BFKL equation to study high--$t$ diffraction in the high energy
limit, which is characterised by a final state with two systems with
transverse momentum $-t$ and far apart in rapidity.
The colour singlet exchange in the non--forward case results in a 
rapidity gap in jet activity. The non--forward BFKL equation thus provides 
a useful theoretical framework to study diffractive physics from first 
principles in QCD.

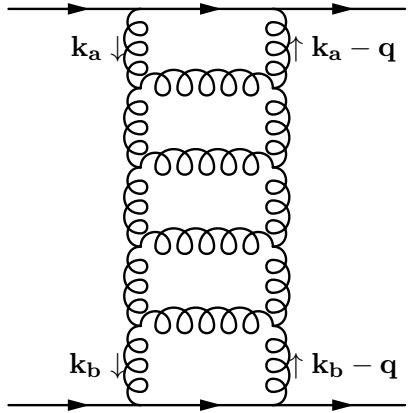
\begin{figure}[tbp]
\begin{fmffile}{qqladder}
  \begin{fmfgraph*}(150,150)
    \fmfset{arrow_len}{3mm} \fmfset{arrow_ang}{15} \fmfpen{1pt} 
    \fmfstraight
    \fmfleft{pi1,pi2} \fmfright{po1,po2}
    \fmf{fermion}{pi1,vup1,vup2,po1} \fmf{fermion}{pi2,vul1,vul2,po2}
    \fmffreeze \fmf{phantom}{vul1,vl14,vl13,vl12,vl11,vup1}
    \fmffreeze \fmf{phantom}{vup2,vl21,vl22,vl23,vl24,vul2} \fmffreeze
    \fmf{gluon}{vl21,vl11}
    \fmf{gluon}{vl22,vl12}
    \fmf{gluon}{vl23,vl13}
    \fmf{gluon}{vl24,vl14}
    \fmf{gluon,label=$\bf{k}_a\downarrow$}{vul1,vl14}
    \fmf{gluon}{vl14,vl13}
    \fmf{gluon}{vl13,vl12}
    \fmf{gluon}{vl12,vl11}
    \fmf{gluon,label=$\bf{k}_b\downarrow$}{vl11,vup1}
    \fmf{gluon,label=$\uparrow \bf{k}_b-\bf{q}$}{vup2,vl21}
    \fmf{gluon}{vl21,vl22}
    \fmf{gluon}{vl22,vl23}
    \fmf{gluon}{vl23,vl24}
    \fmf{gluon,label=$\uparrow \bf{k}_a-\bf{q}$}{vl24,vul2}
  \end{fmfgraph*}
\end{fmffile}\centering
\caption{A Feynman diagram contributing to the LL approximation for 
quark--quark scattering with colour singlet exchange.}
\label{fig:BFKLladder}
\end{figure}

The differential cross--section for diffractive scattering,  
\begin{eqnarray}
\frac{\mathrm{d}\sigma}{\mathrm{d}t} &=& \frac{|A(s,t)|^2}{16\pi s^2}, 
\label{eq:xsec}
\end{eqnarray}
can be described within the BFKL framework by a factorised scattering
amplitude of the form
\begin{eqnarray}
\frac{|A(s,t)|}{s} &=& \left| \int {\mathrm{d}}^2 \mathbf{k}_a\, \int {\mathrm{d}}^2 \mathbf{k}_b 
\, \Phi_{\mathrm{A}} (\mathbf{k}_a,\mathbf{q}) \, \Phi_{\mathrm{B}} (\mathbf{k}_b,\mathbf{q}) \, 
\frac{f \left({\mathbf{k}}_a,{\mathbf{k}}_b,{\mathbf{q}},{\mathrm{Y}} \right)}
{({\mathbf{k}}_a-{\mathbf{q}})^2 {\mathbf{k}}_b^2} \right|,
\end{eqnarray}
where $\mathrm{Y}$ is the rapidity separation of the scattered probes, $\Phi_{\rm
  A}({\bf k}_a,{\bf q})$ and $\Phi_{\rm B}({\bf k}_b,{\bf q})$ are the
process--dependent impact factors and the four--point gluon Green's function,
$f({\bf k}_a,{\bf k}_b,{\bf q},{\rm Y})$, is universal. For illustration, in
Fig.~\ref{fig:BFKLladder} a typical dominant contribution to this colour
singlet exchange in the high energy limit is shown.  As explained below in
Sec.~\ref{Separation} the non--forward BFKL equation describes the evolution
of $f({\bf k}_a,{\bf k}_b,{\bf q},{\rm Y})$ as a function of the rapidity
separation Y. This formalism can be applied, for example, to the study of the
diffractive production of vector mesons in photon--proton collisions at HERA
for large center--of--mass energies and transverse momentum squared
$|t|>\Lambda_{\rm QCD}^2$.  In this kinematical region the Leading Logarithm
(LL) terms $\left(\alpha_s \ln{s/|t|}\right)^n$ generated in the perturbative
series must be resummed. Non--perturbative contributions are included in the
proton parton densities and in the meson light--cone wave function present in
the corresponding impact factor. For the $\rho,~\phi$ and $J/\psi$ mesons the
transverse momentum spectrum and spin density matrix elements have recently
been measured at HERA~\cite{HERA}. From the
theoretical side, there has been an intense activity in the study of these
processes, see e.g.~Ref.~\cite{Largettheory}. The case of photon dissociating 
to a photon has also been studied in
Ref.~\cite{photonphoton}.  Another example of the application
of this approach is the description of events with interjet rapidity gaps in
photon--hadron and hadron--hadron
collisions~\cite{hadronhadron}.

The original analytic solution to the non--forward LL BFKL 
equation~\cite{Lipatov:1985uk}
proved to be significantly more complicated than its forward counterpart. In
order to exhibit the conformal invariance that proved vital in solving the
forward BFKL equation, it is necessary to perform a Fourier transform from
transverse momentum space to impact parameter representation. The non--forward 
BFKL equation at next--to--leading logarithmic (NLL) accuracy 
($\alpha_s \left(\alpha_s \ln{s/|t|}\right)^n$ terms) will be
significantly more complicated, with the conformal invariance being 
explicitly broken by the running of the coupling (as it happens in the 
NLL forward case~\cite{NLLpapers}). It is therefore important 
to investigate new strategies to find the solution to this equation.

In this paper we present a new approach to obtaining the solution to the
non--forward BFKL equation to LL accuracy. The method of solution is based on
the separation of different contributions to the BFKL kernel by a phase space
slice. Once this separation is performed it is possible to apply an iterative
approach similar to the one presented in
Ref.~\cite{Iterative} for the case of the
BFKL equation describing a colour octet exchange. This approach has recently
been generalised~\cite{Solution} to solve this BFKL
equation in the NLL approximation both in QCD
and $N\!=\!4$ supersymmetric Yang--Mills theory~\cite{Andersen:2004uj}
(reviews can be found in~\cite{Reviews}).

In the present work we concentrate on extending this iterative approach to
the solution of the BFKL equation describing diffractive processes. The
presented method of solving the BFKL equation directly in transverse momentum
space has the benefit that it allows for a direct inspection of all involved
momenta. In particular it is possible to study the diffusion of the transverse
scales along the evolution in rapidity.
The same method of solution can be applied to the non--forward BFKL equation
in the NLL approximation.  The corresponding analysis will be performed when
the calculation of the non--forward NLL kernel is completed~\cite{NFNLL}.

The structure of this paper is as follows. In section~\ref{Separation} the LL
BFKL equation in the non--forward case is presented and the phase space
slicing regularisation procedure is performed.  In section~\ref{Solution} the
non--forward LL BFKL equation is solved by iterating the kernel, obtaining in
this way an explicit expression for the four--point gluon Green's function.
In section~\ref{Analysis} the behaviour of the four--point gluon Green's
function is analysed for different values of the momentum transfer. A study
of the diffusion properties is performed of not just the four--point gluon
Green's function, but also, in section~\ref{sec:analysis-toy-cross}, for
cross--sections calculated with toy impact factors. The conclusions
are presented at the end.

\section{The Non--Forward LL BFKL Equation}
\label{Separation}
The LL BFKL equation describing colour singlet exchange with a non--zero
momentum transfer squared, $- t = {\bf q}^2$, was originally calculated in
Ref.~\cite{LLBFKL}. The starting point in the present work will be Eq.~(4.16)
of Ref.~\cite{Forshaw:1997dc} where the integral equation for the Mellin
transform in rapidity of the four--point gluon Green's function, $f_\omega
\left({\bf k}_a,{\bf k}_b, {\bf q}\right)$, is given by
\begin{eqnarray}
\omega  f_\omega \left({\bf k}_a,{\bf k}_b, {\bf q}\right) &=& 
\delta^{(2)} \left({\bf k}_a-{\bf k}_b\right) 
+\frac{{\bar \alpha}_s}{2 \pi} \int d^2{\bf k'} 
\left[\frac{-{\bf q}^2}{\left({\bf k}'-{\bf q}\right)^2 {\bf k}_a^2} 
f_\omega \left({\bf k}',{\bf k}_b, {\bf q}\right) \right. \nonumber\\
&&\hspace{-3.4cm}+
\frac{1}{\left({\bf k}'-{\bf k}_a\right)^2} 
\left(f_\omega \left({\bf k}',{\bf k}_b, {\bf q}\right) -\frac{{\bf k}_a^2}{{{\bf k}'}^2+\left({\bf k}_a-{\bf k}'\right)^2}
f_\omega \left({\bf k}_a,{\bf k}_b, {\bf q}\right) \right)\nonumber\\
&&\hspace{-3.4cm}\left.+\frac{1}{\left({\bf k}'-{\bf k}_a\right)^2} 
\left( 
\frac{\left({\bf k}_a-{\bf q}\right)^2{{\bf k}'}^2}{\left({\bf k}'-{\bf q}\right)^2 {\bf k}_a^2} 
f_\omega \left({\bf k}',{\bf k}_b, {\bf q}\right)
-\frac{\left({\bf k}_a-{\bf q}\right)^2}{\left({\bf k}'-{\bf q}\right)^2 
+\left({\bf k}_a-{\bf k}'\right)^2} 
f_\omega \left({\bf k}_a,{\bf k}_b, {\bf q}\right)\right)\right],
\end{eqnarray}
where $\mathbf{k}_a$ and $\mathbf{k}_b$ describe the two--dimensional
transverse momenta of the exchanged gluons in the $t$--channel (see Fig.~\ref{fig:BFKLladder}) and we define
${\bar \alpha}_s \equiv \alpha_s N_c / \pi$. The driving term $\delta^{(2)}
\left({\bf k}_a-{\bf k}_b\right)$ corresponds to a simple two gluon exchange.
It is more convenient to rearrange the terms in this expression 
and, as the integration variable, to use the transverse momenta of the 
$s$--channel gluons, ${\bf k} = {\bf k}' - {\bf k}_a$, i.e.
\begin{eqnarray}
\omega  f_\omega \left({\bf k}_a,{\bf k}_b, {\bf q}\right) &=& \delta^{(2)} \left({\bf k}_a-{\bf k}_b\right) \nonumber\\
&&\hspace{-4cm}+\frac{{\bar \alpha}_s}{2 \pi} \int d^2{\bf k} 
\left\{ \left[ \frac{1}{{\bf k}^2} 
\left(1 + \frac{\left({\bf k}_a-{\bf q}\right)^2\left({\bf k}+{\bf k}_a\right)^2}{\left({\bf k}+{\bf k}_a -{\bf q}\right)^2 {\bf k}_a^2} \right) 
-\frac{{\bf q}^2}{\left({\bf k}+{\bf k}_a-{\bf q}\right)^2 {\bf k}_a^2} \right]
f_\omega \left({\bf k}+{\bf k}_a,{\bf k}_b, {\bf q}\right) \right.\nonumber\\
&&\hspace{-1.2cm}\left.-\frac{1}{{\bf k}^2} 
\left(\frac{{\bf k}_a^2}{\left({\bf k}+{\bf k}_a\right)^2+{\bf k}^2}
+\frac{\left({\bf k}_a-{\bf q}\right)^2}{\left({\bf k}+{\bf k}_a -{\bf q}\right)^2 + {\bf k}^2} \right)
f_\omega \left({\bf k}_a,{\bf k}_b, {\bf q}\right)\right\}. 
\end{eqnarray}
At this stage it is possible to separate the exchange terms form the 
s--channel contributions by introducing a phase space
slicing parameter $\lambda$ (at NLL it is useful to use dimensional
  regularisation to show the cancellation of infrared divergences, at LL this
  is not needed). It is also convenient to use the following approximation
\begin{eqnarray}
\label{approximation}
f_\omega \left({\bf k}+{\bf k}_a,{\bf k}_b, {\bf q}\right) &=&
f_\omega \left({\bf k}+{\bf k}_a,{\bf k}_b, {\bf q}\right) 
\left(\theta\left({\bf k}^2 - \lambda^2\right)+
\theta\left(\lambda^2-{\bf k}^2\right)\right) \nonumber\\
&& \hspace{-1cm}
\simeq f_\omega \left({\bf k}+{\bf k}_a,{\bf k}_b, {\bf q}\right) 
\theta\left({\bf k}^2 - \lambda^2\right) +
f_\omega \left({\bf k}_a,{\bf k}_b, {\bf q}\right)
\theta\left(\lambda^2-{\bf k}^2\right).
\end{eqnarray}
In all the results presented in this paper we have made sure this
approximation is valid by checking that the four--point Green's function is
insensitive to the value of the slicing parameter for small values of
$\lambda$.

Therefore, the non--forward BFKL equation for the Green's function can 
be written in a very simple form:
\begin{eqnarray}
\label{SimpleEqn}
\left(\omega -\omega_0 \left({\bf k}_a, {\bf q}, \lambda \right)
\right) f_\omega \left({\bf k}_a,{\bf k}_b, {\bf q}\right) ~=~ 
\delta^{(2)} \left({\bf k}_a-{\bf k}_b\right) \nonumber\\
&&\hspace{-5cm}+ \int \frac{d^2{\bf k}}{\pi {\bf k}^2} 
\theta \left({\bf k}^2-\lambda^2\right)
\xi \left({\bf k}_a,{\bf k},{\bf q}\right)
f_\omega \left({\bf k}+{\bf k}_a,{\bf k}_b, {\bf q}\right). 
\end{eqnarray}
In order to write the equation in this way the following notation has been
introduced:
\begin{eqnarray}
\xi\left({\bf k}_a,{\bf k},{\bf q}\right) &=&
\frac{{\bar \alpha}_s}{2}   
\left( 1 + \frac{\left({\bf k}_a-{\bf q}\right)^2\left({\bf k}+{\bf k}_a\right)^2-{\bf q}^2 {\bf k}^2}{\left({\bf k}+{\bf k}_a -{\bf q}\right)^2 {\bf k}_a^2}
\right),
\end{eqnarray}
and 
\begin{eqnarray}
\omega_0 \left({\bf k}_a, {\bf q},\lambda\right) &=& 
\frac{{\bar \alpha}_s}{2 \pi} \int 
\frac{d^2{\bf k}}{{\bf k}^2} \left[\theta \left(\lambda^2-{\bf k}^2\right)
\left(1 + \frac{\left({\bf k}_a-{\bf q}\right)^2\left({\bf k}+{\bf k}_a\right)^2-{\bf q}^2 {\bf k}^2}{\left({\bf k}+{\bf k}_a -{\bf q}\right)^2 {\bf k}_a^2}
\right) \right.\nonumber\\
&-&\left.\frac{{\bf k}_a^2}{\left({\bf k}+{\bf k}_a\right)^2+{\bf k}^2}
-\frac{\left({\bf k}_a-{\bf q}\right)^2}{\left({\bf k}+{\bf k}_a -{\bf q}\right)^2 + {\bf k}^2} \right].
\end{eqnarray}
The latter expression corresponds to the Regge trajectory in our
regularisation. In the case of $\mathbf{q}={\bf 0}$, the trajectory for the
colour--octet exchange should be obtained, and indeed we find
\begin{eqnarray}
\label{trajforw}
\omega_0 \left({\bf k}_a, {\bf 0},\lambda\right) &=& 
\frac{{\bar \alpha}_s}{\pi} \int 
\frac{d^2{\bf k}}{{\bf k}^2} \left[\theta \left(\lambda^2-{\bf k}^2\right)-
\frac{{\bf k}_a^2}{\left({\bf k}+{\bf k}_a\right)^2+{\bf k}^2}\right]
~=~ - {\bar \alpha}_s \ln{\frac{{\bf k}_a^2}{\lambda^2}}.
\end{eqnarray}
Moreover, the whole solution to the non--forward BFKL equation has the correct
limit for $q\to0$ as found in Ref.~\cite{Iterative}. 

The expression for the non--forward LL Regge trajectory can be simplified if
we make use of the forward limit in Eq.~(\ref{trajforw}), i.e.
\begin{eqnarray}
\omega_0 \left({\bf k}_a, {\bf q},\lambda\right) &=&
\frac{1}{2} \, \left( \omega_0 \left({\bf k}_a, {\bf 0},\lambda\right)+
\omega_0 \left({\bf k}_a-{\bf q}, {\bf 0},\lambda\right) \right) \nonumber \\
&+&\frac{{\bar \alpha}_s}{2 \pi} \int 
\frac{d^2{\bf k}}{{\bf k}^2} \, \theta \left(\lambda^2-{\bf k}^2\right)
\left(\frac{\left({\bf k}_a-{\bf q}\right)^2\left({\bf k}+{\bf k}_a\right)^2-{\bf q}^2 {\bf k}^2}{\left({\bf k}+{\bf k}_a -{\bf q}\right)^2 {\bf k}_a^2}-1
\right). 
\end{eqnarray}
The last integral is negligible when $\lambda$ is small, and
therefore the non--forward trajectory takes the simple form
\begin{eqnarray}
\label{eq:trajappr}
\omega_0 \left({\bf k}_a, {\bf q},\lambda\right) &\simeq&
-\frac{{\bar \alpha}_s}{2} \left(\ln{\frac{{\bf k}_a^2}{\lambda^2}}
+\ln{\frac{\left({\bf k}_a-{\bf q}\right)^2}{\lambda^2}}\right).
\end{eqnarray}
With these conventions we proceed in the next section to iterate
Eq.~(\ref{SimpleEqn}) to find the solution for the non--forward
four--point gluon Green's function.

\section{Solution to the Equation}
\label{Solution}

The non--forward LL BFKL Green's function is the solution to the integral
equation
\begin{eqnarray}
f_\omega \left({\bf k}_a,{\bf k}_b, {\bf q}\right) ~=~ 
\frac{1}{\omega -\omega_0 \left({\bf k}_a, {\bf q}, \lambda \right)}
\left\{\delta^{(2)} \left({\bf k}_a-{\bf k}_b\right) \right.\nonumber\\
&&\hspace{-6cm}\left.+ \int \frac{d^2{\bf k}}{\pi {\bf k}^2} 
\theta \left({\bf k}^2-\lambda^2\right)
\xi \left({\bf k}_a,{\bf k},{\bf q}\right)
f_\omega \left({\bf k}+{\bf k}_a,{\bf k}_b, {\bf q}\right) \right\}. 
\label{one_over_w}
\end{eqnarray}
for $\lambda\to0$. If this expression is iterated, the Green's function can
be expressed in terms of a kernel per iteration acting on the initial
condition with a series of poles in the $\omega$ complex plane:
\begin{eqnarray}
\label{iterating}
f_{\omega} \left({\bf k}_a ,{\bf k}_b,{\bf q}\right) &=& 
\frac{\delta^{(2)} \left({\bf k}_a - {\bf k}_b \right)}{\omega - {\omega}_0 \left({\bf k}_a,{\bf q},\lambda\right)} \nonumber\\
&&\hspace{-1cm}+ \int \frac{d^2 {\bf k}_1}{\pi {\bf k}_1^2} 
\frac{\theta\left({\bf k}_1^2-\lambda^2\right)\, \xi\left({\bf k}_a,{\bf k}_1,{\bf q}\right)}{\omega - {\omega}_0 \left({\bf k}_a,{\bf q},\lambda\right)}
\frac{\delta^{(2)} \left({\bf k}_a +{\bf k}_1 - {\bf k}_b\right)}
{\omega - {\omega}_0 \left({\bf k}_a+{\bf k}_1,{\bf q},\lambda\right)} \nonumber\\
&&\hspace{-1cm}+ \int \frac{d^2 {\bf k}_1}{\pi {\bf k}_1^2}
\int \frac{d^2 {\bf k}_2}{\pi {\bf k}_2^2} 
\frac{\theta\left({\bf k}_1^2-\lambda^2\right)\, \xi\left({\bf k}_a,{\bf k}_1,{\bf q}\right)}{\omega - {\omega}_0 \left({\bf k}_a,{\bf q},\lambda\right)}
\frac{\theta\left({\bf k}_2^2-\lambda^2\right)\, \xi\left({\bf k}_a+{\bf k}_1,{\bf k}_2,{\bf q}\right)}{\omega - {\omega}_0 \left({\bf k}_a+{\bf k}_1,{\bf q},\lambda\right)}\nonumber\\
&&\hspace{1cm}\times\frac{\delta^{(2)} \left({\bf k}_a +{\bf k}_1 +{\bf k}_2- {\bf k}_b\right)}{\omega - {\omega}_0 \left({\bf k}_a+{\bf k}_1+{\bf k}_2,{\bf q},\lambda\right)} \nonumber\\
&&\hspace{-1cm}+ \cdots 
\end{eqnarray}
Each action of the kernel corresponds to an interaction between the reggeised
gluons exchanged in the $t$--channel, building up, in this way, the LL BFKL
ladder. The poles are integrated over when going from the $\omega$--plane to
rapidity space using the Mellin transform
\begin{eqnarray}
f \left({\bf k}_a,{\bf k}_b, {\bf q}, {\rm Y}\right) 
&=& \frac{1}{2 \pi i}
\int_{a-i \infty}^{a+i \infty} d\omega ~ e^{\omega {\rm Y}} f_{\omega} 
\left({\bf k}_a ,{\bf k}_b, {\bf q}\right),
\end{eqnarray}
where Y is the rapidity span of the BFKL ladder.

This integration can be performed to finally obtain the solution of the
non--forward LL BFKL equation for the four--point Green's function
as (with $y_0 \equiv$ Y)
\begin{eqnarray}
\label{eq:nonf_soln}
&&\hspace{-1cm}f({\bf k}_a ,{\bf k}_b, {\bf q}, {\rm Y}) 
= \left(
\frac{\lambda^2}{{\bf k}_a^2}\frac{\lambda^2}{\left({\bf k}_a-{\bf q}\right)^2}\right)^{\frac{\bar{\alpha}_s}{2} \, {\rm Y}}
\left\{  \frac{}{} \delta^{(2)} ({\bf k}_a - {\bf k}_b) \right. \\
&+& \sum_{n=1}^{\infty} \prod_{i=1}^{n} \int d^2{\bf k}_i 
\frac{\theta\left({\bf k}_i^2- \lambda^2\right)}{\pi {\bf k}_i^2} \, 
\xi \left({\bf k}_a+\sum_{l=1}^{i-1} {\bf k}_l,{\bf k}_i,{\bf q}\right)
\nonumber \\
&\times& \left. \int_0^{y_{i-1}} d y_i 
\left(\frac{\left({\bf k}_a+\sum_{l=1}^{i-1} {\bf k}_l\right)^2}
{\left({\bf k}_a+\sum_{l=1}^i {\bf k}_l\right)^2}
\frac{\left({\bf k}_a+\sum_{l=1}^{i-1} {\bf k}_l -{\bf q}\right)^2}
{\left({\bf k}_a+\sum_{l=1}^i {\bf k}_l-{\bf q}\right)^2}\right)
^{\frac{\bar{\alpha}_s}{2} \, y_i} \delta^{(2)} \left(\sum_{l=1}^{n}{\bf k}_l 
+ {\bf k}_a - {\bf k}_b \right) \right\}. \nonumber
\end{eqnarray}
This solution has the correct forward limit of Ref.~\cite{Iterative} 
when the momentum transfer tends to zero. We would like to stress again 
that this method of solution is directly
applicable to the non--forward BFKL equation also at NLL accuracy, with the
difference that the introduction of the phase space slice is more conveniently 
performed in the language of dimensional regularisation.  

\section{Analysis of the Gluon Green's Function}
\label{Analysis}
To study the dependence of the non--forward LL BFKL four--point gluon Green's
function on the transverse momentum scales we choose to integrate over all
external angles and define the quantity
\begin{eqnarray}
{\bar f} \left(\left|{\bf k}_a\right|,\left|{\bf k}_b\right|,\left|{\bf q}\right|, {\rm Y}\right) &=& \int_0^{2 \pi} d \theta_{qa} \int_0^{2 \pi} d \theta_{qb} \, f \left({\bf k}_a, {\bf k}_b, {\bf q}, {\rm Y}\right), 
\end{eqnarray}
where $\theta_{qi}$ is the angle between the vectors ${\bf k}_i$ and ${\bf q}$.

\begin{figure}
\centerline{\epsfig{file=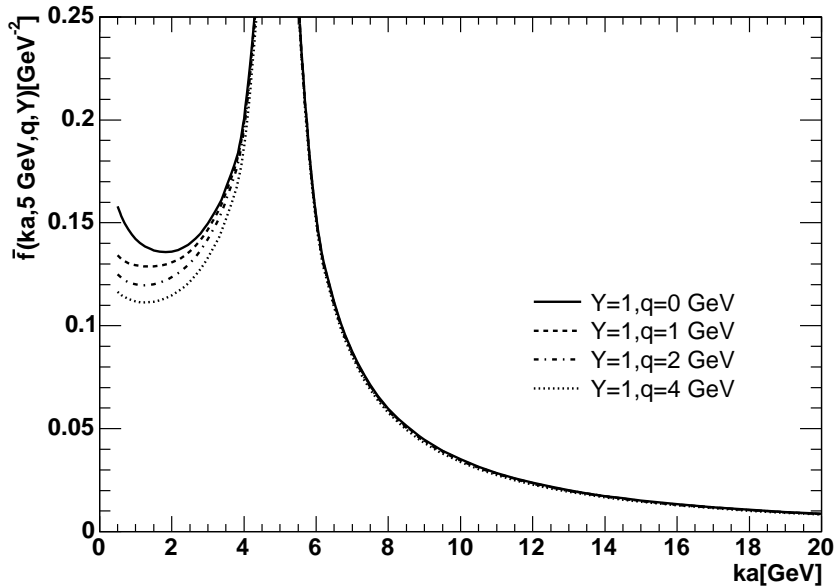,width=12cm,angle=0}}
\caption{The angular integrated Green's function for fixed 
$\left|{\bf k}_b\right|=5$ GeV as 
a function of $\left|{\bf k}_a\right|$ for different values of the momentum
transfer. The rapidity span is low, Y = 1.} 
\label{2}
\end{figure}

\begin{figure}
\centerline{\epsfig{file=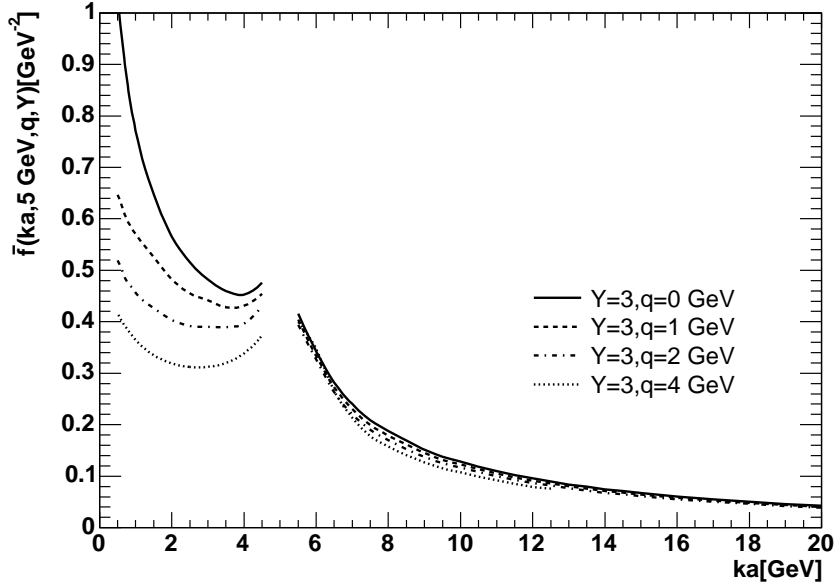,width=12cm,angle=0}}
\caption{The angular integrated Green's function for fixed 
$\left|{\bf k}_b\right|=5$ GeV as 
a function of $\left|{\bf k}_a\right|$ for different values of the momentum
transfer $q$. The rapidity span is Y = 3.} 
\label{3}
\end{figure}

As a first analysis, in Fig.~\ref{2} the value of $\left|{\bf k}_b\right|$ is
fixed to 5 GeV and the dependence on $\left|{\bf k}_a\right|$ is studied.
When the value of the modulus of both momenta coincides, the angular
integrated Green's function shows a $\delta$--functional behaviour corresponding to the
two gluon exchange limit. This dependence is caused by the driving term of
the integral equation whose influence is stronger for lower energies. When
Y is increased from 1 in Fig.~\ref{2} to ${\rm Y}=3$ in Fig.~\ref{3} the
influence of the driving term diminishes as a consequence of a larger number
of effective rungs in the BFKL ladder. It is interesting to note that the
influence of the momentum transfer $q$ is larger in regions of low scales
of $k_a$.  The general trend is that the four--gluon Green's function
decreases with increasing $q$.  Comparing Fig.~\ref{2} with Fig.~\ref{3} it
can be seen that this effect is increasing with rapidity.

Let us now proceed to the study of the diffusion of the transverse scales in
the BFKL ladder with increasing rapidity span. We choose in this analysis to
concentrate on the evolution of the transverse scale along the left hand side
of the ladder depicted on Fig.~\ref{fig:BFKLladder}, but, as previously
mentioned, one of the benefits of having the solution to the non--forward
BFKL equation expressed in terms of explicit phase space integrals as in
Eq.~(\ref{eq:nonf_soln}) is that it is possible to study any of the momenta
in the ladder.

\begin{figure}
  \begin{center}
\epsfig{file=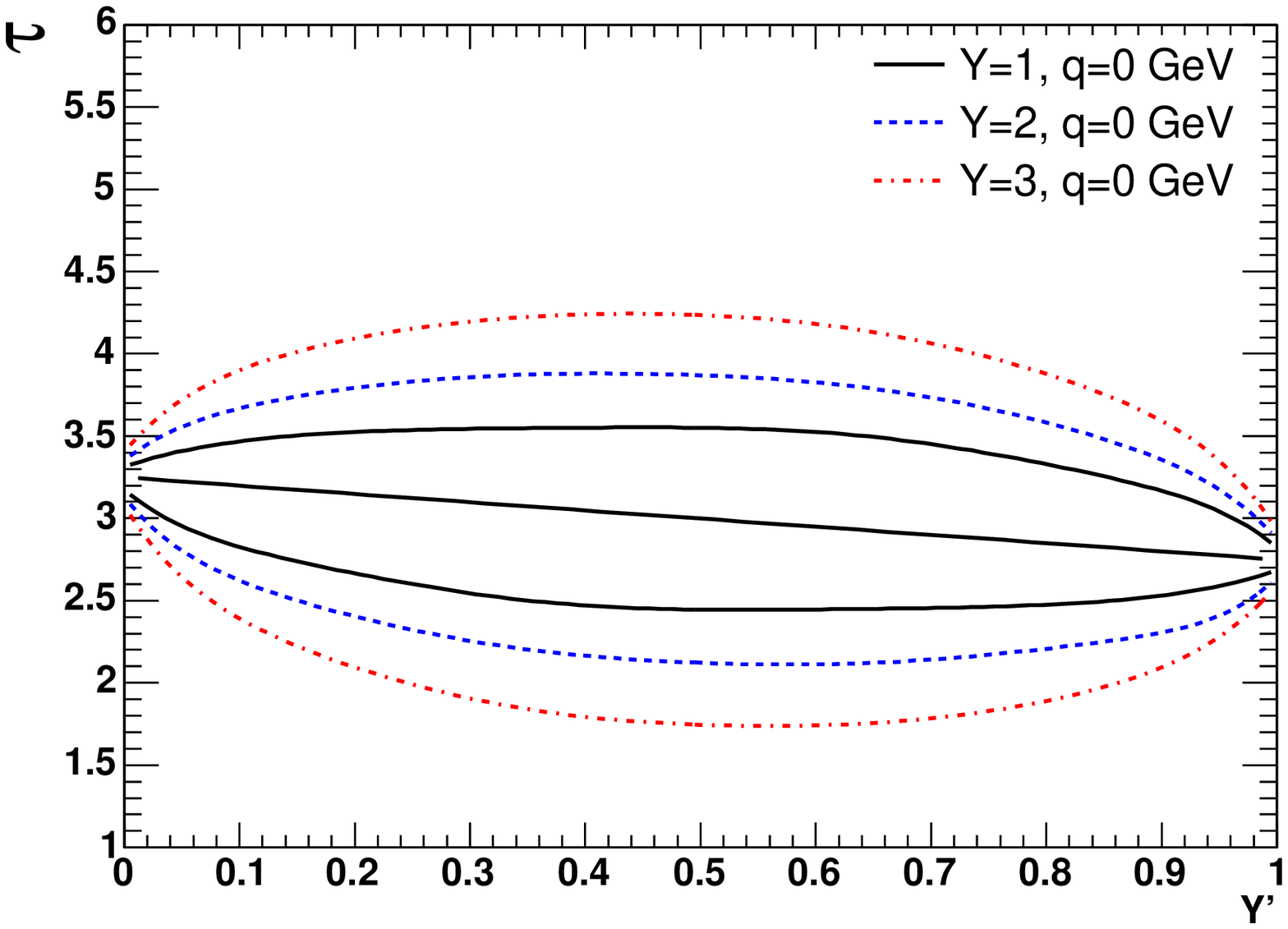,width=7.5cm,angle=0}
\epsfig{file=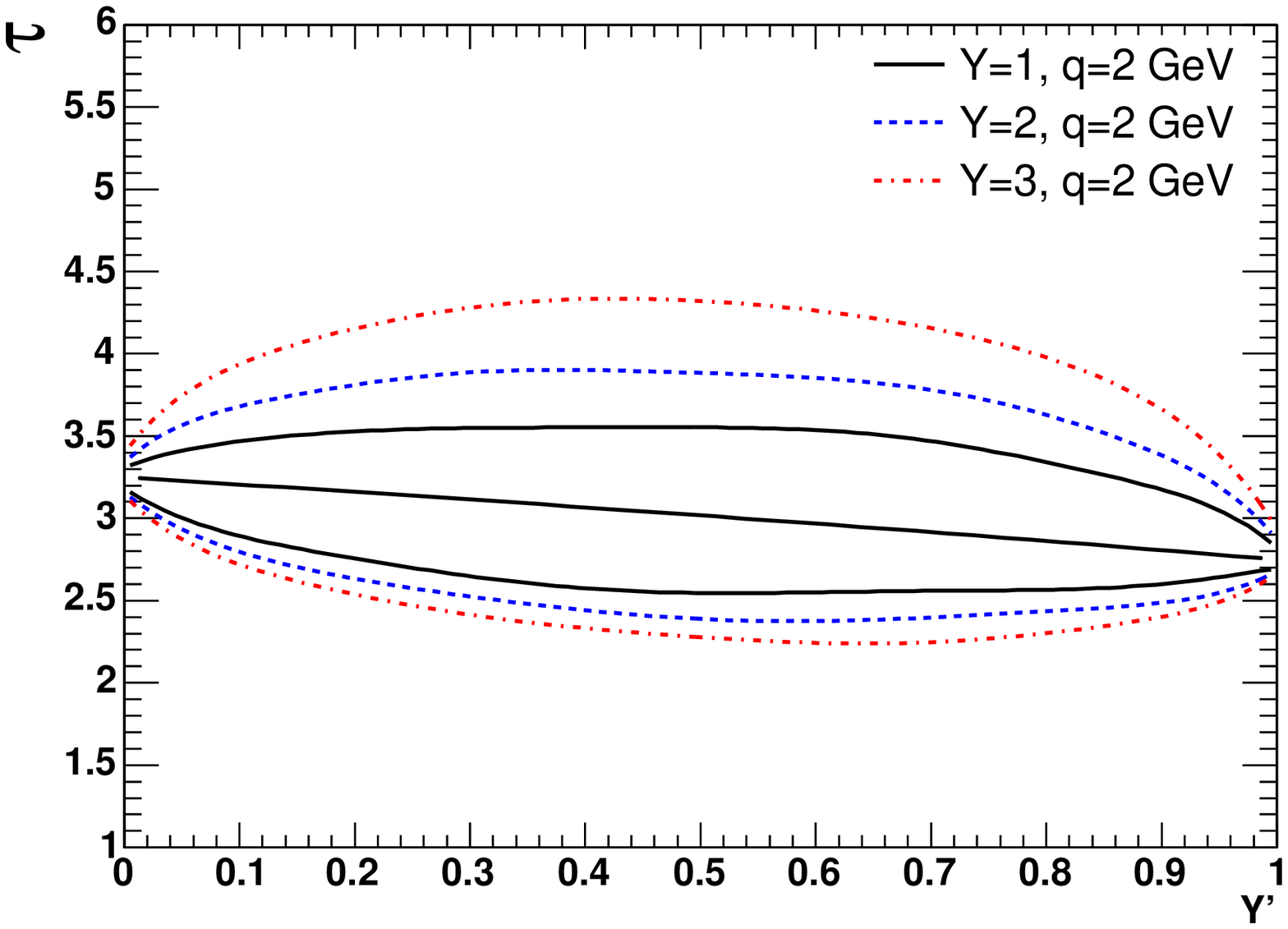,width=7.5cm,angle=0}
\end{center}
\caption{The diffusion properties of the gluon Green's function in terms of
  the lines formed by $\langle\tau\rangle$, $\langle\tau\rangle+\sigma_1$,
  and $\langle\tau\rangle-\sigma_2$ along the BFKL ladder. Shown for
  $k_a=5$~GeV, $k_b=4$~GeV, $q=0$~GeV (left) and $q=2$~GeV (right), for
  rapidity spans of ${\rm Y}=1,2,3$.}
\label{fig:diff_q}
\end{figure}

Diffusion is normally studied in terms of the mean of the transverse momentum
along the ladder, with the width of the distribution indicated by the
standard deviation (see
e.g.~Ref.~\cite{Forshaw:1997dc,DIF}). This choice
is useful, since it allows analytic studies. However, a plot of the mean
internal transverse momentum plus/minus the standard deviation as a function
of the rapidity along the BFKL ladder fails, by construction, to display the
different behaviour of diffusion to low and high scales. Given the
possibilities offered by the solution written in terms of explicit phase
space integrals, we therefore choose to study diffusion in terms of the
average value $\langle\tau\rangle$ of $\tau=\ln((k_a+\sum
k_i)^2/\mathrm{GeV^2})$ as a function of the rapidity ${\rm Y}'$ along the 
ladder.
Specifically, for a given value of $k_a, k_b$, and Y we solve the
non--forward BFKL equation according to Eq.~(\ref{eq:nonf_soln}) by a
MonteCarlo integration method. For each configuration point in $n-$momenta
phase space $\{k_i,y_i\}$ we can trace the evolution of $\tau$ along the ladder
and, at the same time, calculate the weight of this configuration to the total
solution. In this way it is possible to calculate both the average value of
$\tau$ along the ladder, $\langle\tau\rangle{\scriptstyle( {\rm Y}')}$, and the
quantities
\begin{align}
  \begin{split}
    \sigma^2_1{\scriptstyle( {\rm Y}')}&= \frac{\displaystyle 2\int_{<\tau>{\scriptstyle( {\rm Y}')}}^\infty d\tau (\tau-\langle\tau\rangle{\scriptstyle( {\rm Y}')})^2 \bar{f}(k_a,k_b,q,{\rm Y})}{\displaystyle\int_0^\infty d\tau \bar{f}(k_a,k_b,q,{\rm Y})},\\
    \sigma^2_2{\scriptstyle( {\rm Y}')}&=\frac{\displaystyle 2\int_0^{<\tau>{\scriptstyle( {\rm Y}')}} d\tau
    (\tau-\langle\tau\rangle{\scriptstyle( {\rm Y}')})^2 \bar{f}(k_a,k_b,q,{\rm Y})}{\displaystyle\int_0^\infty d\tau \bar{f}(k_a,k_b,q,{\rm Y})}.
\end{split}
\end{align}
For a gluon Green's function $f(\mathbf{k}_a,\mathbf{k}_b,\mathbf{q},{\rm Y})$ that
is symmetric in $\tau$, the lines of $\langle\tau\rangle{\scriptstyle( {\rm Y}')}$,
$\langle\tau\rangle{\scriptstyle( {\rm Y}')}+\sigma_1{\scriptstyle( {\rm Y}')}$, and
$\langle\tau\rangle{\scriptstyle( {\rm Y}')}-\sigma_2{\scriptstyle( {\rm Y}')}$ would
reproduce the plot of the mean plus/minus the standard deviation. This is
true for $q=0$~GeV, as shown in Fig.~\ref{fig:diff_q}. Here we have plotted
the above mentioned three lines for $k_a=5$~GeV, $k_b=4$~GeV, $q=0$~GeV,
$\alpha_s=0.23$, and ${\rm Y}=1,2,3$. ${\rm Y}'$ is rescaled to lie between 0 and 1, so as to plot all the three cases on the same figure.

However, for $q>0$~GeV the distribution of $\tau$ is no longer symmetric, as
can be seen on the right hand side plot in Fig.~\ref{fig:diff_q}, where we
have plotted the same quantities for $q=2$~GeV. It is apparent that there is
less diffusion to smaller scales, the average is unchanged, and the diffusion
to scales larger than the average is almost unchanged compared to the case
$q=0$~GeV.

It is also apparent that for the set of parameters investigated here, there
is no significant diffusion into regions where the coupling is expected to
become unperturbatively large. As it is well known, the influence of softer 
scales is larger at higher energies, but in this work we confirm the fact 
that the diffusion into the infrared is drastically reduced when there is 
some momentum transfer, acting, in this way, as an efficient infrared 
cut--off. It will be interesting to investigate if this picture holds at NLL. 

\section{Analysis of a Toy Cross--Section} 
\label{sec:analysis-toy-cross}
\begin{figure}[tbp]
  \centering
  \epsfig{file=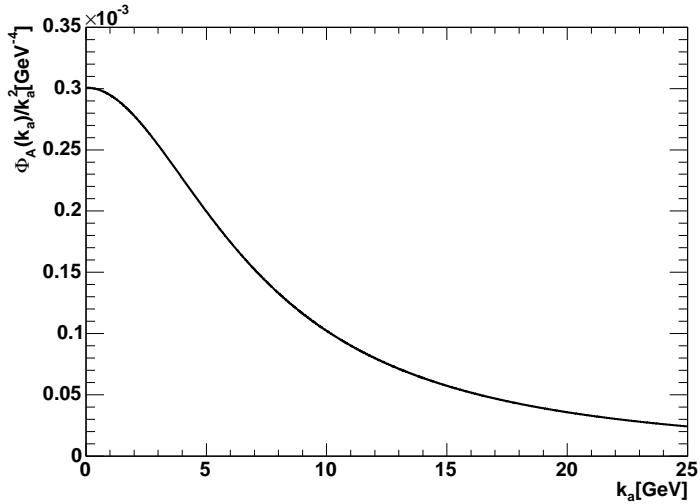,width=10cm}
  \caption{The impact factor $\Phi_A(k_a)/k_a^2$ for $q=0$~GeV and $m=3.1$~GeV.}
  \label{fig:imp_fac}
\end{figure}

Let us now turn to the study of diffractive cross--sections. To this end we
need to define some suitable impact factors. We choose a generic example from
Ref.\cite{Forshaw:1997dc} modelling the photo--production of a vector meson:
\begin{eqnarray}
  \Phi_{\rm A}(\mathbf{k}_a,\mathbf{q})=\alpha_s h^2 \int\mathrm{d}\rho \, 
\mathrm{d}\tau \rho \,
  (1-\rho)\left[\frac 1 {\left(\mathbf{q}^2 \rho^2 \tau(1-\tau)+m^2\right)} 
- \frac 1
{\left((\mathbf{k}_a-\rho \, \mathbf{q})^2\tau(1-\tau)+m^2 \right)}\right],
\end{eqnarray}
and similarly for $\Phi_{\rm B}$. Here, $h$ is a normalisation constant which
we choose arbitrarily such that $\alpha_s h^2=1$. This choice obviously means
that the normalisation of the toy cross--section reported in this paper is
completely arbitrary. We have plotted this impact factor divided by $k_a^2$ in
Fig.~\ref{fig:imp_fac} for $q=0$~GeV and the mass of the meson $m=3.1$~GeV.
One further complication arises compared to studies of the four--point gluon
Green's function due to the integration over $k_a$ and $k_b$ when calculating 
the differential cross--section as in Eq.~(\ref{eq:xsec}). The
approximations of Eqs.~(\ref{approximation}) and~(\ref{eq:trajappr}) are valid
only when $\lambda\ll k_a$. When $k_a$ is integrated over it is therefore
not possible to have $\lambda$ fixed. This situation is similar to the one
encountered in Ref.\cite{Andersen:2004nm}. This problem can be solved in
several different ways. In the present analysis we choose to always have
$\lambda<20 \, k_a$ and $\lambda\le1$~GeV. We have checked that the results 
here presented do not depend on these choices.

\begin{figure}
\centerline{\epsfig{file=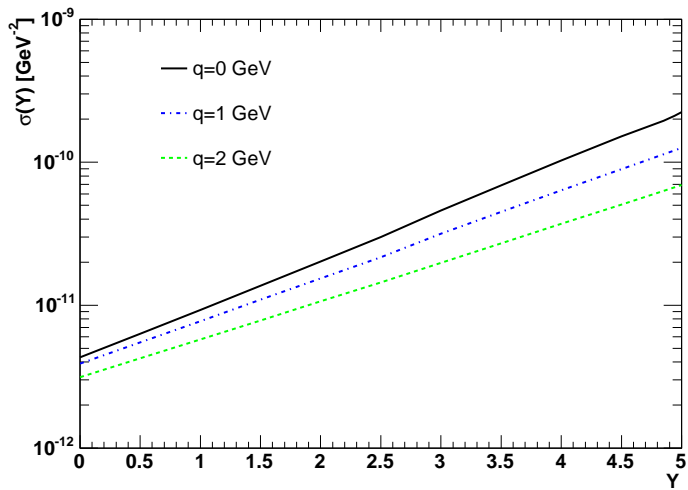,width=10cm}}
\caption{The toy cross--section as a function of the rapidity separation of the
  produced vector mesons ($m=3.1$~GeV), each produced with transverse momentum $q=0,1,2$~GeV.}
\label{fig:xsecvsdeltay}
\end{figure}

On Fig.~\ref{fig:xsecvsdeltay} we have plotted the resulting cross--section
of Eq.~(\ref{eq:xsec}) as a function of the rapidity span of the BFKL
ladder for $q=0,1,2$~GeV, $\alpha_s=0.2$ and $m=3.1$~GeV. The exponential
rise of the cross--section as a function of rapidity is evident for all $q$. 
In the previous section we showed how the Green's function diminishes as 
the momentum transfer rises, this translates here into smaller cross--sections. 

\begin{figure}
  \begin{center}
    \epsfig{file=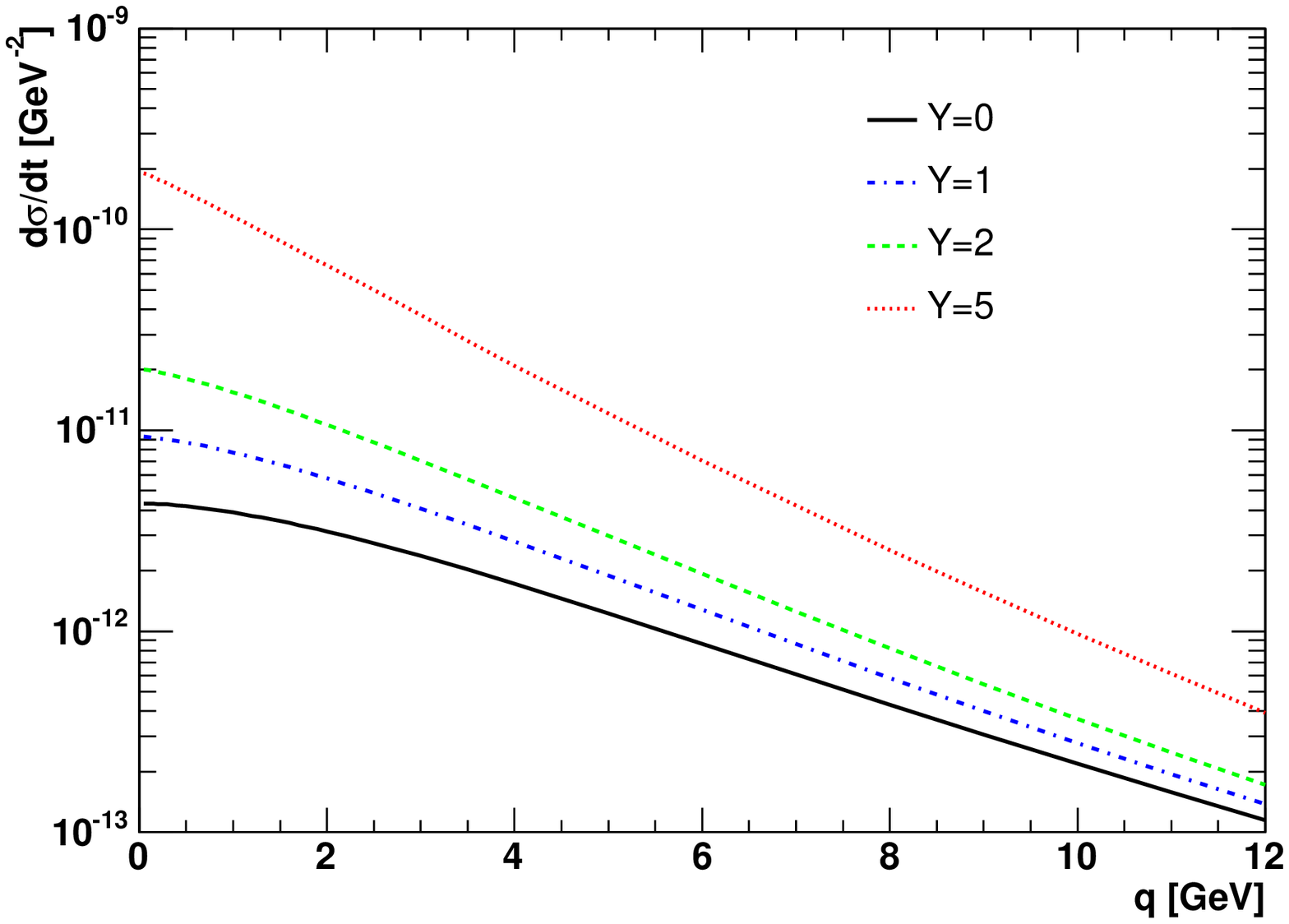,width=7.5cm}
    \epsfig{file=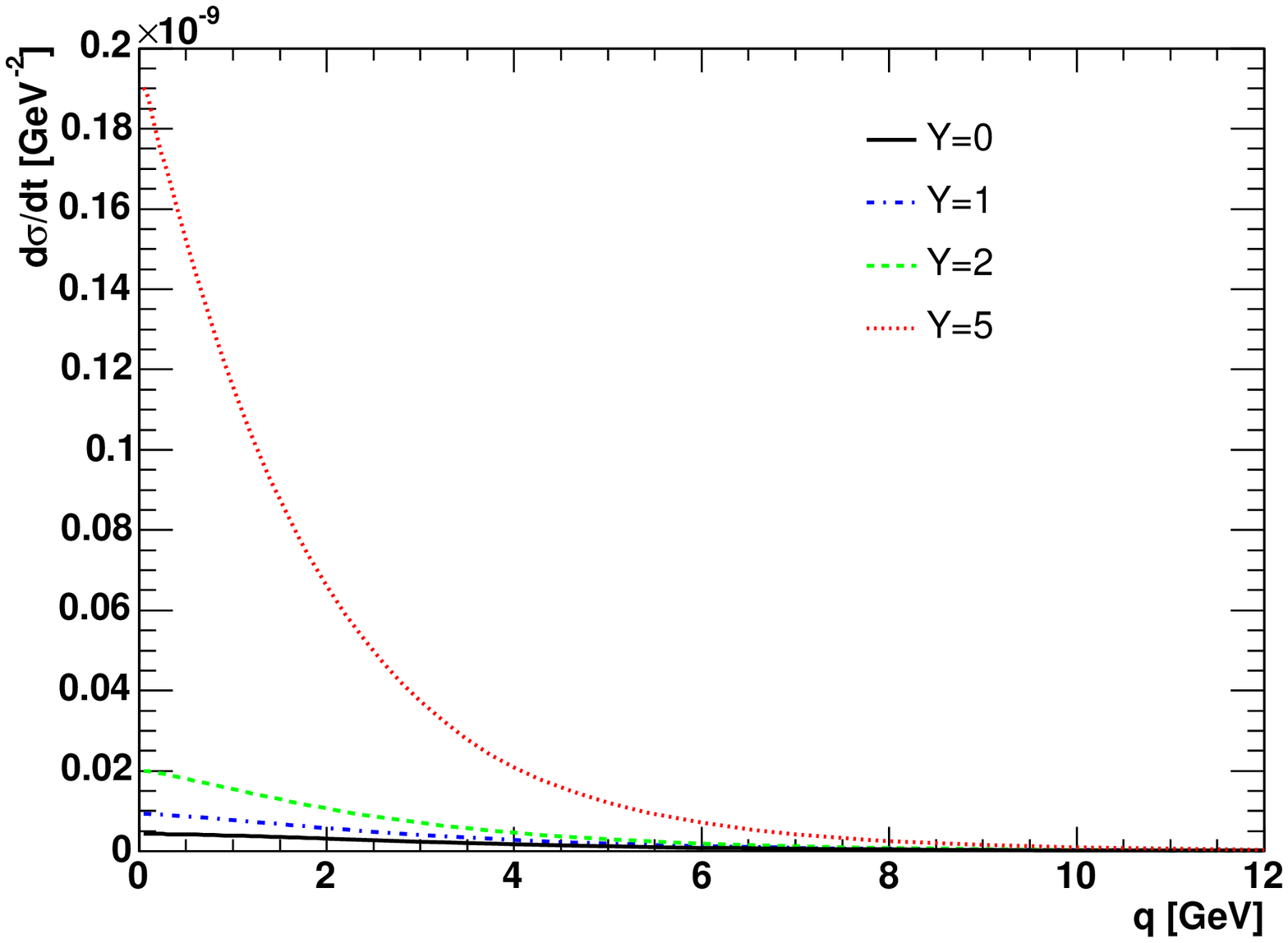,width=7.5cm}
  \end{center}
\caption{The toy cross--section as a function of the transverse momentum of
    the produced vector mesons, for rapidity separations of ${\rm Y}=0,1,2,5$ units
    of rapidity.} 
\label{fig:xsecvsq}
\end{figure}

In Fig.~\ref{fig:xsecvsq} we have plotted the $q-$dependence of the 
cross--section for rapidity spans of ${\rm Y}=0,1,2,5$. We see an exponential 
fall--off with $q$ for all Y. Furthermore, we observe that
$\mathrm{d}\sigma/\mathrm{d}t$ at $t=0$~$\mathrm{GeV}^2$ increases with Y. 
The curves in Fig.~\ref{fig:xsecvsq} are presented on a linear (right) and 
a logarithmic (left) scale.

These results could have also been obtained using an analytic approach to
solving the non--forward BFKL equation at LL accuracy. Nevertheless it is when 
discussing diffusion properties that the method presented in this paper has
advantages. For the analysis of diffusion in cross--sections, a second issue
arises compared to the study of diffusion of the Green's function. The
perturbative scale is now set by both the meson mass $m$ and the momentum 
transfer squared $-t$, while the scales $k_a$ and $k_b$ are no longer fixed. It
therefore becomes interesting to study not only the distribution of the
average transverse scale $\langle\tau\rangle$ along the ladder, but also at
the ends of it, where it will describe the average scale of the transverse 
momentum connecting the impact factors to the gluon Green's function.

\begin{figure}
  \begin{center}
\epsfig{file=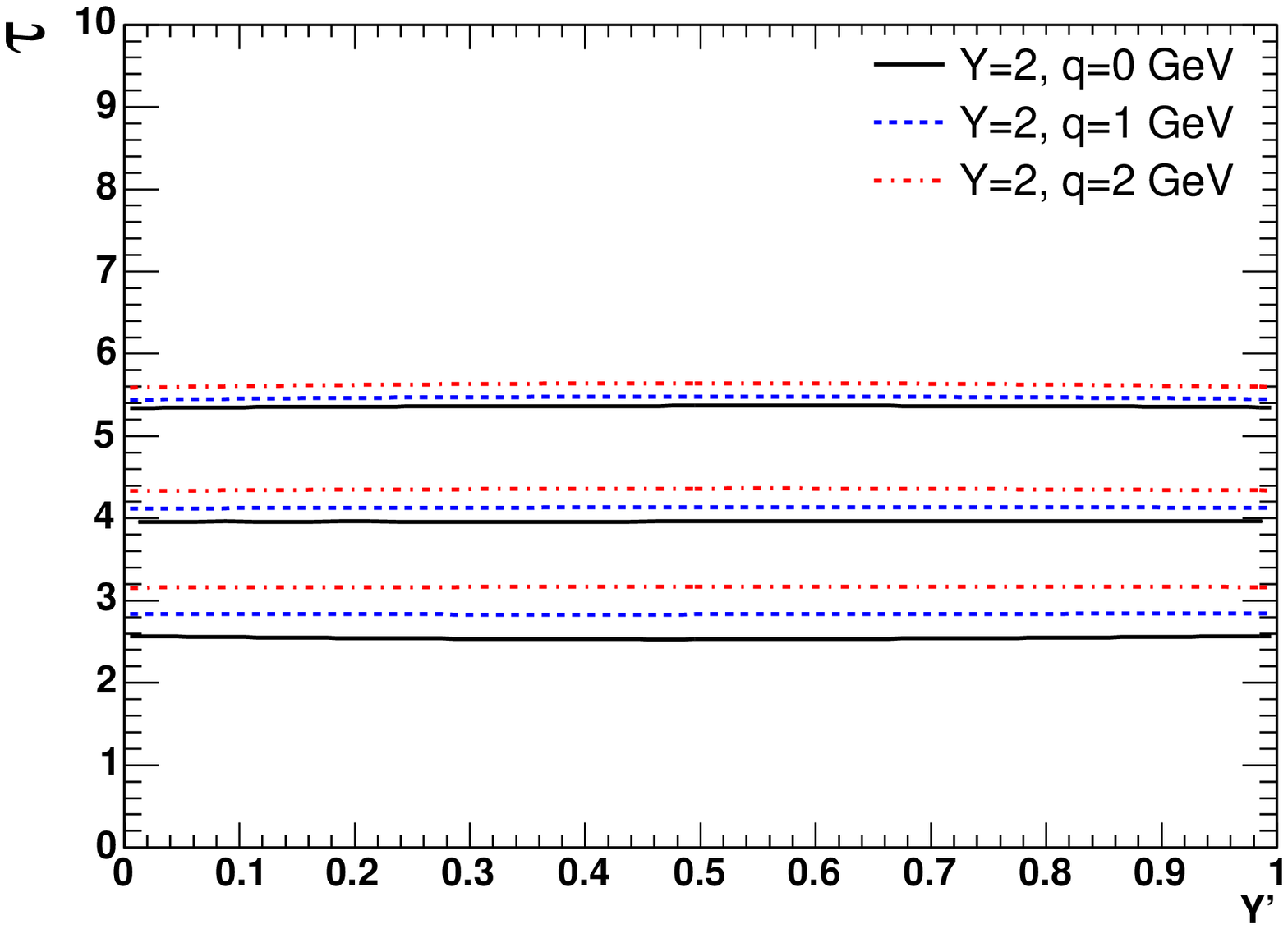,width=7.5cm,angle=0}
\epsfig{file=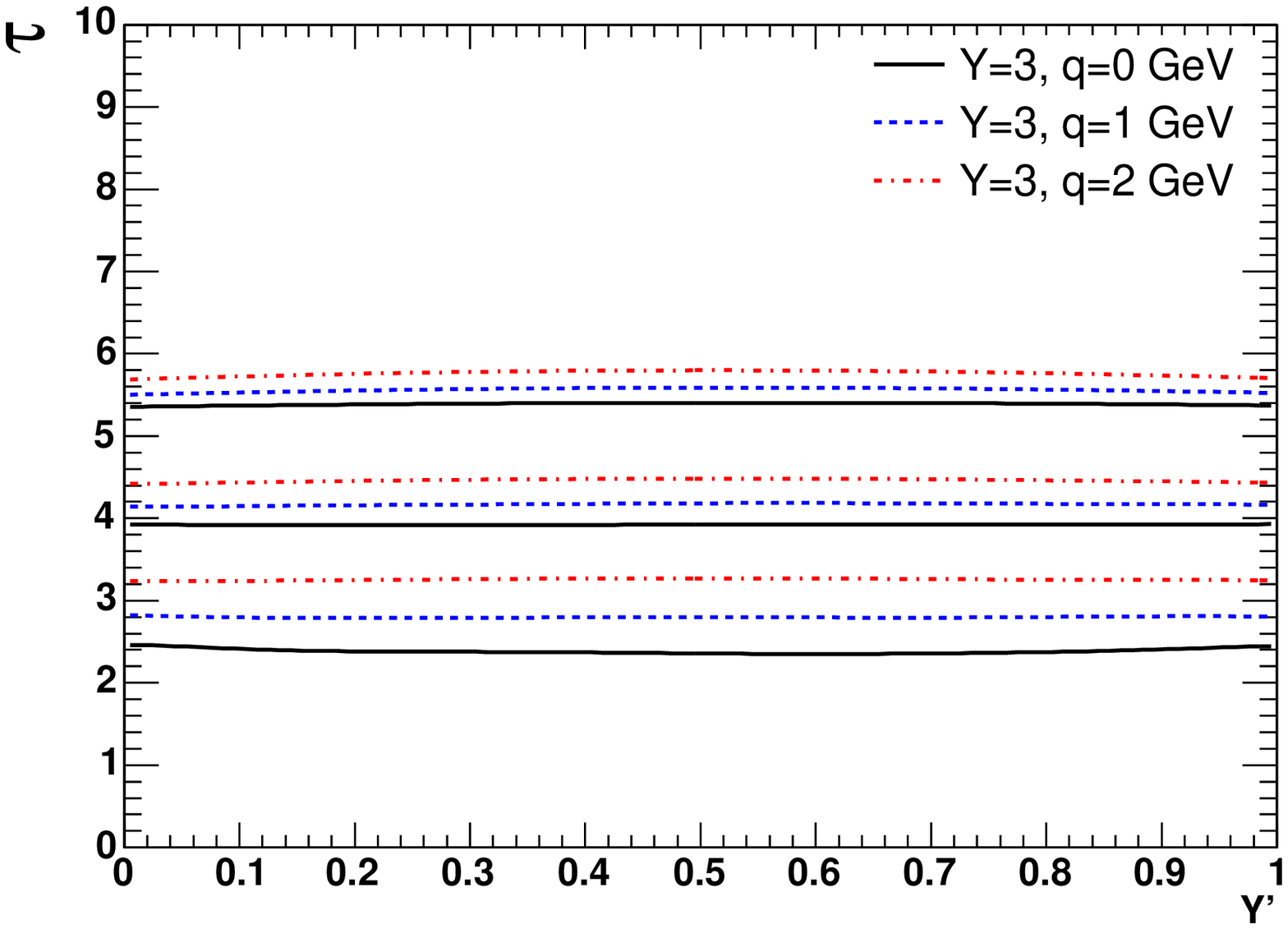,width=7.5cm,angle=0}
\end{center}
\caption{The range of relevant internal transverse momentum scales in terms
  of $\langle\tau\rangle, \langle\tau\rangle+\sigma_1,
  \langle\tau\rangle-\sigma_2$ for the diffractive cross-- section for
  $\alpha_s=0.2$, $m=3.1$~GeV, ${\rm Y}=2,3$, and $q=0,1,2$~GeV. ${\rm Y}'$ is 
the rescaled (to unity) rapidity along the BFKL ladder.}
\label{fig:xsec_diff}
\end{figure}

In Fig.~\ref{fig:xsec_diff} we have plotted the distribution of the average
momentum scale in a similar way to the one used in Fig.~\ref{fig:diff_q} for
the Green's function. It is interesting first to note that the spread along
the chain is not significantly larger than that at the ends of the chain (for
the rapidity spans considered here). Secondly, it is comforting to see that
the typical scales remain perturbative for all values of the momentum
transfer $q$, for the chosen value of the mass of the vector meson
($m=3.1$~GeV). The average value of $\tau$ for $q=0$~GeV
($\langle\tau\rangle\approx3.9$) corresponds to a scale of the internal momenta of
the BFKL exchange of roughly 7~GeV. The logarithmic scale of momenta on
Fig.~\ref{fig:diff_q} and Fig.~\ref{fig:xsec_diff} emphasises the region of
soft momenta. It might therefore be helpful to study the effect of the
increase in scattering momenta on the internal scales directly: an increase
of $q$ from 0~GeV to 2~GeV leads to and increase in the upper (UV) lines
corresponding to an increase in the internal momenta of roughly 2.4~GeV, while
the increase in the lower (IR) lines corresponds to an increase of roughly
1.7~GeV.

\section{Conclusions}
We have presented a new solution to the non--forward BFKL equation at leading
logarithmic accuracy that allows a study of diffusion properties directly in
momentum space. We have investigated the behaviour of the gluon Green's 
function as a function of transverse scales, including a study of the 
IR/UV diffusion. Then we extended this study to the analysis of a 
toy cross--section. 

The presented framework is very efficient for solving BFKL evolution
equations and the solution allows immediate insight into the momentum
configurations of the evolution. We hope to extend the iterative method for
solving the non--forward BFKL equation to next--to--leading logarithmic
accuracy, once the appropriate integral kernel is calculated.

\subsubsection*{Acknowledgements}
\label{sec:acknowledgements}
We acknowledge useful discussions with Jochen
Bartels, Victor Fadin, Jeff Forshaw, Lev Lipatov and Leszek Motyka. ASV 
would like to thank the Cavendish Laboratory at the University of Cambridge 
and both authors wish to thank the CERN Theory Division for hospitality. JRA
acknowledges the support of PPARC (postdoctoral fellowship PPA/P/S/2003/00281).

\end{document}